\documentclass{article}

\usepackage{microtype}
\usepackage{graphicx}
\usepackage{subfigure}
\usepackage{bbold}
\usepackage{booktabs} 

\usepackage{hyperref}

\usepackage[accepted]{icml2020}

\icmltitlerunning{Generative Modelling for Controllable Audio Synthesis of 
Expressive Piano Performance}

\usepackage{amsmath}

\newcommand{\LB}[1] {\mathcal{L}(\theta, \phi; #1)}

\newcommand{\z}[1]{\mathbf{z}}

\newcommand{\cat}[1]{\mathbf{c}}

\newcommand{\E}[2] {\mathbb{E}_{#1}[ #2 ]}
\newcommand{\Dsymbol}[1] {\mathcal{D}_{\mathrm{#1}}}
\newcommand{\KLD}[2] {\Dsymbol{KL}\big( #1 \| #2 \big)}

\newcommand{\vh}[2]{\textbf{#1}^{\scriptsize{\textrm{#2}}}}
\begin{document}
\setlength{\abovedisplayskip}{1pt}
\setlength{\belowdisplayskip}{1pt}
\twocolumn[
\icmltitle{Generative Modelling for Controllable Audio Synthesis of \\ Expressive Piano Performance}



\icmlsetsymbol{equal}{*}

\begin{icmlauthorlist}
\icmlauthor{Hao Hao Tan}{equal,to}
\icmlauthor{Yin-Jyun Luo}{equal,to,goo}
\icmlauthor{Dorien Herremans}{to,goo}
\end{icmlauthorlist}

\icmlaffiliation{to}{Singapore University of Technology and Design}
\icmlaffiliation{goo}{Institute of High Performance Computing, A*STAR, Singapore}

\icmlcorrespondingauthor{Hao Hao Tan}{haohao\_tan@sutd.edu.sg}

\icmlkeywords{Machine Learning, ICML}

\vskip 0.3in
]



\printAffiliationsAndNotice{\icmlEqualContribution} 

\begin{abstract}
We present a controllable neural audio synthesizer based on Gaussian Mixture Variational Autoencoders (GM-VAE), which can generate realistic piano performances in the audio domain 
that closely follows temporal conditions of two essential style features for piano performances: \textit{articulation} and \textit{dynamics}. 
We demonstrate how the model is able to apply fine-grained style morphing over the course of synthesizing the audio.
This is based on conditions which are latent variables that can be sampled from the prior or inferred from other pieces.
One of the envisioned use cases is to inspire creative and brand new interpretations for existing pieces of piano music.

\end{abstract}
\vspace{-20pt}
\section{Introduction}
Synthesizing audio of piano performances from MIDI 
requires either a huge collection of recordings of individual notes, or a model that simulates an actual piano. These approaches, however, have two main limitations: (i) the ``stitching" of individual notes might not optimally capture 
the various interactions between notes~\cite{hawthorne2018enabling}; and (ii) the quality of the synthesized audio is restricted by the recordings in the sound library or the piano simulator. This further motivates the effort to build realistic piano synthesizers. 
In this work, we propose a neural network based synthesizer which takes the onset 
roll as input, 
and generates realistic piano performances in the audio domain. 
Our model takes the onset roll instead of the complete piano roll as input, thereby waiving the necessity of fine-grained frame and velocity information, which are often unavailable,
to synthesize expressive piano performances. Without the constraint of frame and velocity information, performance style transfer could also be achieved by interpreting each onset note with the style inferred from a given audio performance.
On top of the onset 
roll, the generation is further conditioned on variables corresponding to style features, which enable the rendition of expressive piano performances.
In particular, we consider \textit{articulation} and \textit{dynamics}, which are two significant features related to expressiveness in piano performances, and base our framework on variational autoencoders (VAEs)~\cite{kingma2013auto} with a Gaussian mixture prior~\cite{jiang2016variational}.

The underlying task of this work is to build a neural audio synthesizer that maps piano performances in MIDI 
to audio.
The current literature either focuses on such direct mapping without additional controllability with regard to 
performance styles~\cite{hawthorne2018enabling, manzelli2018combining}, or concerns only the global attributes such as synthesizing different instruments~\cite{kim2019neural}.
We distinguish ourselves from the previous works by incorporating GM-VAE~\cite{jiang2016variational}, which has been applied to speech~\cite{ hsu2018hierarchical} and instrument modelling ~\cite{luo2019}, to disentangle two significant expressive performance factors. This allows us to achieve creative applications such as gradual style morphing over time. 
The sequence of conditions can be 
sampled from the prior or inferred from other pieces, as shown in Section \ref{sec:result}.
\vspace{-8pt}
\section{Experimental Details}

\begin{figure}
  \includegraphics[width=\columnwidth]{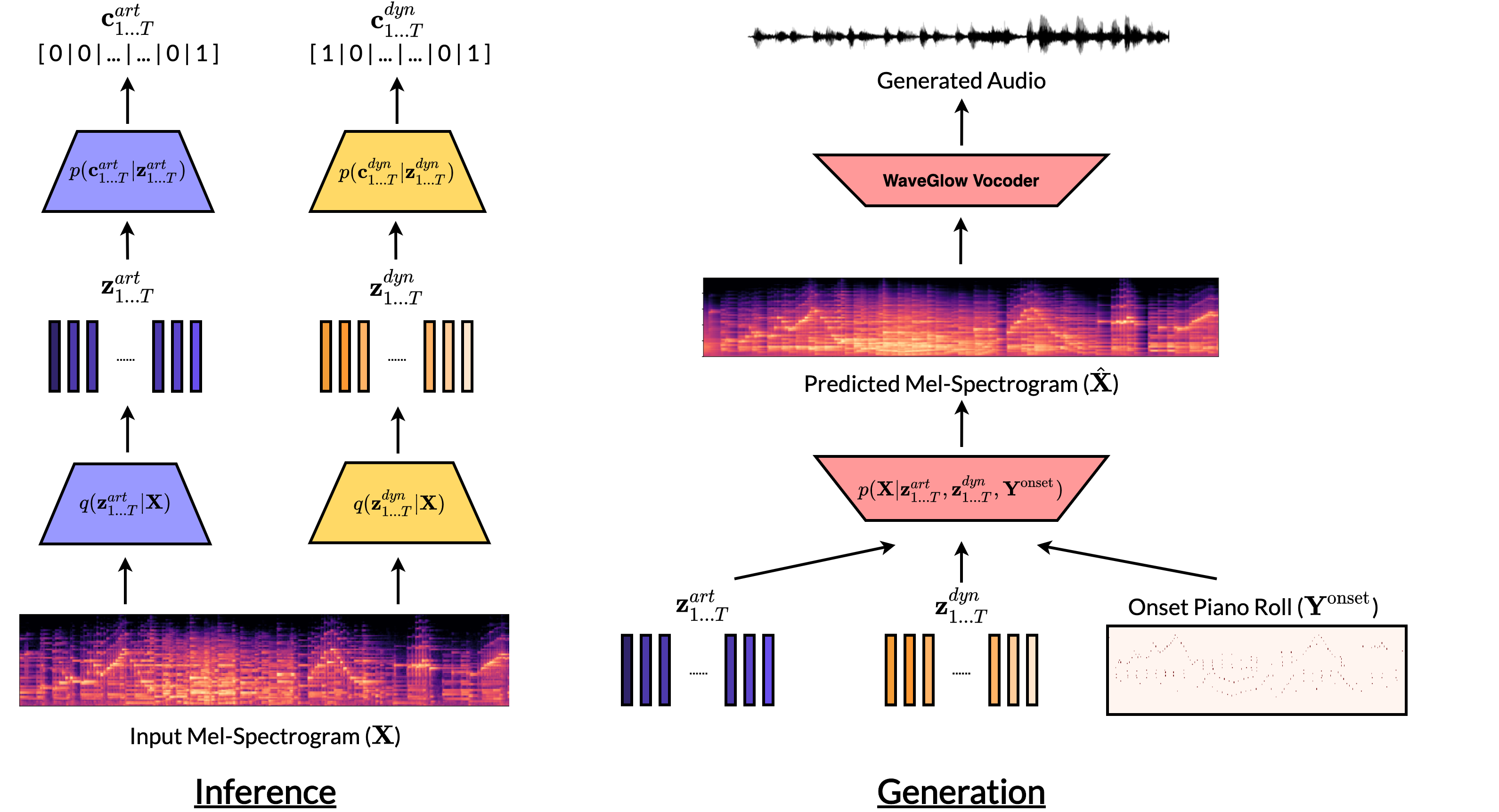}
  \vspace{-15pt}
  \caption{Model architecture.}
  \label{fig:model}
  \vspace{-15pt}
\end{figure}

\textbf{Data Representation}: We use the MAESTRO v2.0.0 dataset \cite{hawthorne2018enabling} which consists of 1,282 performances with aligned audio-MIDI pairs, 
and we split the data
as annotated.
We train on 20-second random crops from audio clips,
which are converted into a log-scale Mel-spectrogram with 80 Mel-filters ($\vh{X}{}$). 
The MIDI note sequences are represented as piano rolls,
from which the onset roll (\(\textbf{Y}^{\scriptsize{\textrm{onset}}}\)) is extracted.
We represent both articulation (\textit{staccato}, \textit{legato}) and dynamics (loud, soft) as binary sequences \(\textbf{c}^{\scriptsize{\textrm{art}}}_{1...T}\),  \(\textbf{c}^{\scriptsize{\textrm{dyn}}}_{1...T} \in \{0, 1\}\), each of which corresponds to two components of a Gaussian mixture.
\(c^{\scriptsize{\textrm{art}}}_{t} = 1\) if at least one note is held at \(t\); 
and \(c^{\scriptsize{\textrm{dyn}}}_{t} = 1\) if the average velocity across all notes is greater than 70 at \(t\), 
which the threshold is determined by our preliminary data analysis.

\textbf{Model Formulation}: Figure \ref{fig:model} shows our model architecture, which is adapted from~\cite{hsu2018hierarchical}.
The joint distribution 
$p_\theta(\vh{X}{}, \vh{z}{art}, \vh{z}{dyn}|\vh{Y}{onset}, \vh{c}{art}, \vh{c}{dyn}) = p_\theta(\vh{X}{}|\vh{Y}{onset}, \vh{z}{art}, \vh{z}{dyn}) p_\theta(\vh{z}{art}|\vh{c}{art})
p_\theta(\vh{z}{dyn}|\vh{c}{dyn})$, 
where $\theta$ refers to parameters of the generation network.
Both the prior distributions $p_\theta(\vh{z}{art})$ and $p_\theta(\vh{z}{dyn})$ are Gaussian mixtures of two components.
A variational distribution $q_\phi(\vh{z}{}|\vh{X}{})$ is introduced to approximate the true posterior,
where $\phi$ refers to parameters of the inference network.
The model is trained to optimize the evidence lower bound:
\vspace{-10pt}

\begin{equation}
\begin{split}
    \LB{\vh{X}{}} &= \E{q_{\phi}(\vh{z}{art}|\vh{X}{})
    q_{\phi}(\vh{z}{dyn}|\vh{X}{})}{\log p_{\theta}(\vh{X}{}|\vh{Y}{onset},\vh{z}{art}, \vh{z}{dyn})} \\
    - &\KLD{q_{\phi}(\vh{z}{art}|\vh{X}{})}{p(\vh{z}{art}|\vh{c}{art})} \\
    - &\KLD{q_{\phi}(\vh{z}{dyn}|\vh{X}{})}{p(\vh{z}{dyn}|\vh{c}{dyn})}
\end{split}
\end{equation}
Both the generation and inference network are implemented with two-layer bidirectional LSTMs\footnote{Source code: \href{https://github.com/gudgud96/piano-synthesis}{https://github.com/gudgud96/piano-synthesis}}.
Note that we simplify \(\textbf{c}^{\scriptsize{\textrm{art}}}_{1...T}\) and  \(\textbf{c}^{\scriptsize{\textrm{dyn}}}_{1...T}\) as \(\textbf{c}^{\scriptsize{\textrm{art}}}\) and  \(\textbf{c}^{\scriptsize{\textrm{dyn}}}\) (similarly for the latent variables $\vh{z}{art}$ and $\vh{z}{dyn}$).
For each of $\vh{c}{art}$ and $\vh{c}{dyn}$, an additional cross-entropy loss is introduced such that the posterior $p(\vh{c}{}|\vh{z}{})$ can also learn from the labelled ground-truth $\vh{c}{}$.

\textbf{Audio Synthesis}: We leverage WaveGlow~\cite{prenger2019waveglow} to invert the Mel-spectrogram to audio, due to its fast inference and superior performance~\cite{govalkar2019comparison, zhao2020transferring}.
We adopt the implementation from~\cite{memwaveglow}. 

\section{Results and Discussion}\label{sec:result}

\begin{figure}
  \includegraphics[width=\columnwidth]{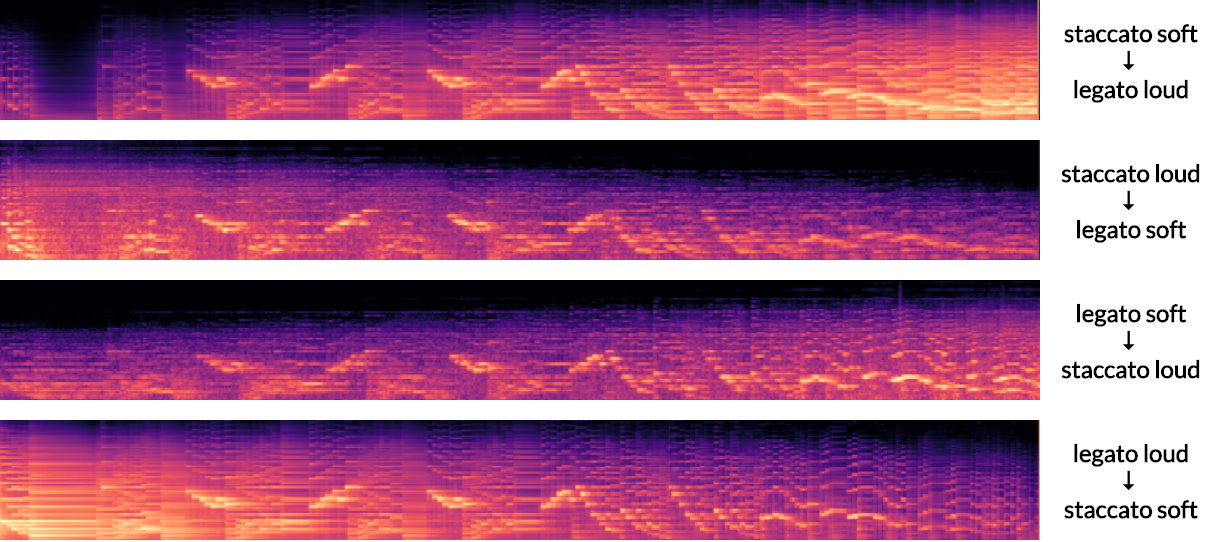}
  \vspace{-20pt}
  \caption{Generated Mel-spectrograms of a given piece, with combinations of gradual morphing of articulation and dynamics.}
  \vspace{-8pt}
  \label{fig:spec-output}
\end{figure}

\textbf{Gradual Style Morphing Over Time}: Utilizing the Gaussian mixture prior distribution enables style morphing by linear interpolation between mixture components. In particular, given $\mu_{0}^{\textrm{art}}$ and $\mu_{1}^{\textrm{art}}$ representing the mean vectors of mixture components corresponding to \textit{staccato} and \textit{legato}, 
we can set $\vh{z}{art}_t = \mu_{0}^{\textrm{art}} + (\mu_{1}^{\textrm{art}} - \mu_{0}^{\textrm{art}}) \times \frac{t}{T}$ in the latent space of articulation (similarly for $\vh{z}{dyn}_t$). 
In other words, the sequence of
conditioning vectors travel from $\mu_{0}^{\textrm{art}}$ to $\mu_{1}^{\textrm{art}}$, whereby we can expect the articulation of the synthesized piano performance to morph gradually from \textit{staccato} to \textit{legato}.
Figure \ref{fig:spec-output} demonstrates the generated Mel-spectrograms of four different scenarios.
From \textit{staccato} to \textit{legato}, one can observe that notes are gradually sustained longer; 
and from \textit{soft} to \textit{loud}, the amplitude increases over time and more mid-high frequencies are covered, which results in a higher level of perceived energy in the human auditory system  \cite{fletcher1933loudness}.

\begin{figure}
  \includegraphics[width=\columnwidth]{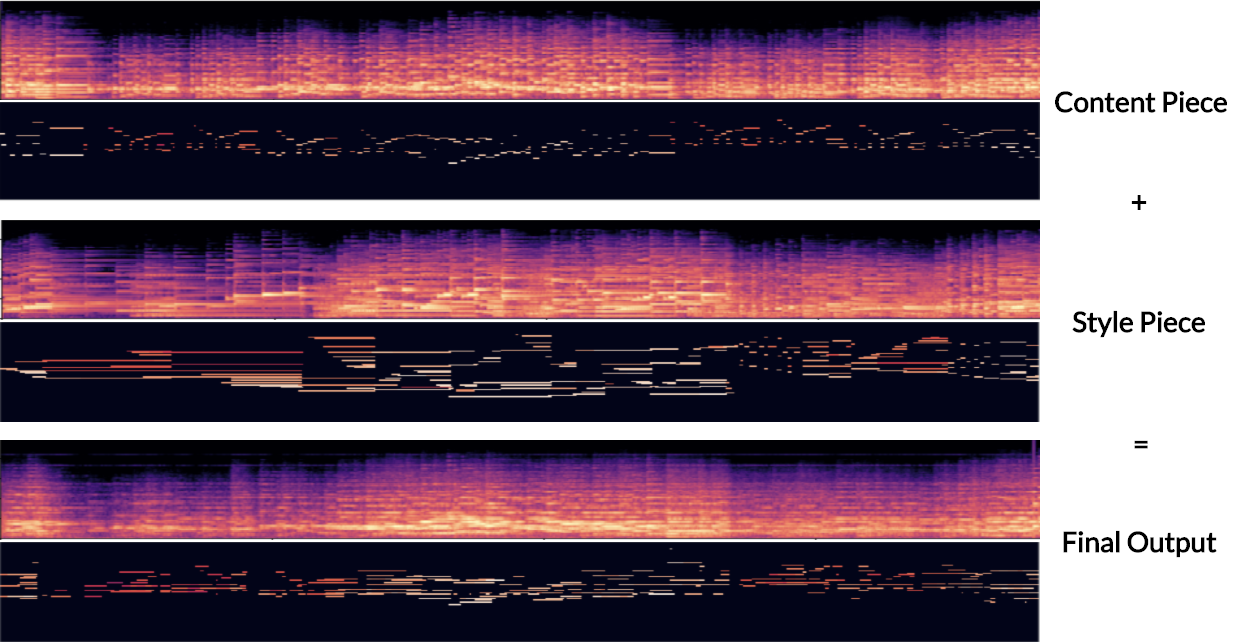}
  \vspace{-20pt}
  \caption{An example of performance style transfer. The piano roll for the final output is estimated using the state-of-the-art model for piano transcription
  \cite{hawthorne2017onsets}.}
  \label{fig:spec-transfer}
  \vspace{-10pt}
\end{figure}

\textbf{Performance Style Transfer}: 
In addition to sampling from the prior distribution, we can also infer the sequence of conditioning vectors from another piece of music. 
Specifically, we let $\vh{z}{art} \sim q_{\phi}(\vh{z}{art} | \vh{X}{style})$ (similarly for $\vh{z}{dyn}$), where $\vh{X}{style}$ is the \textit{style piece} that would determine the fine-grained style over time of the synthesized piano performance.
Figure \ref{fig:spec-transfer} shows an example that renders a piece from the Baroque era (which is more detached and constant in terms of dynamics) in the style of a piece from the Romantic era (which is more \textit{legato} and expressive in terms of dynamics). 
By observing both the Mel-spectrograms and piano rolls, one can see that the final output closely follows the style features of the style piece in terms of note duration (articulation) and amplitude (dynamics), 
while preserving the musical content. Audio examples can be found online.\footnote{ \href{https://piano-performance-synthesis.github.io}{https://piano-performance-synthesis.github.io}}

We envision that this framework could learn to achieve fine-grained control on multiple performance style factors, 
which 
allows us to explore 
new performance directions for any given piece, even by taking inspirations from other pieces via style transfer. Future work will involve extending the set of performance features (e.g. onset deviation, pedalling) 
in order to generate more realistic piano performance.
\newpage
\section*{Acknowledgements}

We would like to thank the anonymous reviewers for their constructive reviews. This work is supported by MOE Tier 2 grant no. MOE2018-T2-2-161, SRG ISTD 2017 129, and Singapore International Graduate Award (SINGA) provided by the Agency for Science, Technology and Research (A*STAR), under reference number SING-2018-01-1270.
\bibliography{example_paper}
\bibliographystyle{icml2020}



\end{document}